\documentclass[floatfix,preprint, onecolumn, showpacs,nofootinbib]{revtex4-1}
\usepackage{natbib}
\usepackage{times}
\usepackage{amssymb,amsbsy,amsmath,amsfonts}
\usepackage{graphicx}
\usepackage{float}
\usepackage{morefloats}
\usepackage{rotating}
\usepackage{srcltx}
\usepackage{slashed}
\usepackage{subfigure}
\begin{document}

\title{Finite-volume effects on octet-baryon masses in covariant baryon chiral perturbation theory}

\author{ Li-sheng Geng,$^{1,2}$ Xiu-lei Ren$^1$, J. Martin-Camalich,$^3$ and W. Weise$^2$}
\affiliation{
$^1$School of Physics and Nuclear Energy Engineering, Beihang University,  Beijing 100191,  China\\
$^2$Physik Department, Technische Universit\" at M\"unchen, D-85747 Garching, Germany\\
$^3$Department of Physics and Astronomy, University of Sussex, BN1 9Qh,
    Brighton, UK.}

 \begin{abstract}
We study finite-volume effects on the masses of the ground-state octet baryons using covariant baryon chiral perturbation theory (ChPT) up to next-to-leading order by analyzing
the latest $n_f=2+1$ lattice Quantum ChromoDynamics (LQCD) results from the NPLQCD collaboration. Contributions of virtual decuplet baryons are taken into
account using the ``consistent'' coupling scheme. We compare our results with those obtained from heavy baryon ChPT and show that, although both approaches can describe well the lattice data, the underlying physics is different: In HBChPT, virtual decuplet baryons play a more important role than they do in covariant ChPT. This is because the virtual octet baryon contributions to  finite-volume corrections  are larger in covariant ChPT than   in HBChPT, while the contributions of intermediate decuplet baryons are smaller, because of  relativistic effects.
 We observe that for the octet baryon masses, at fixed $m_\pi L$ ($\gg1$) finite-volume corrections decrease as $m_\pi$ approaches its physical value, provided that the strange quark mass is at or close to its physical value, as in most LQCD setups.

\end{abstract}


\date{\today}

\maketitle

\section{Introduction}
Understanding the origin of the masses of light hadrons has long been a central topic in strong-interaction physics. Due to the non-perturbative nature of the strong interaction at low energies, calculations based on first principles have only become possible with the advent of lattice Quantum ChromoDynamics (LQCD).
LQCD has made remarkable progress in studies of strong-interaction physics in the past decade (see, e.g., Refs.~\cite{Bazavov:2009bb,Hagler:2009ni}). Nowadays, fully dynamical calculations have become standard and therefore one of the most-difficult-to-estimate uncertainties related to ``quenching" effects in LQCD calculations of early times have been removed. Nonetheless, LQCD simulations still have to adopt unphysical simulation parameters: larger than physical light quark masses $m_{u/d}$, finite lattice volume $V=T L^3$, finite lattice spacing $a$, etc.  To obtain physical results, extrapolations to the physical point in terms of $m_{u/d}$, $L$ ($T$), and $a$ must be performed,\footnote{In this work, we limit our discussions to LQCD simulations of zero-temperature physics} i.e., $m_{u/d}\rightarrow m_{u/d}(\mathrm{phys.})$, $L (T)\rightarrow\infty$, and $a\rightarrow0$.  

The extrapolation in light quark masses $m_{u/d}$ is usually termed as ``chiral extrapolation" [see Refs.~\cite{  Leinweber:2003dg,Bernard:2003rp,Procura:2003ig,Bernard:2005fy} for some early studies of the nucleon mass in SU(2)].  In the real world, chiral symmetry and its breaking pattern govern the dynamics of low-energy strong interaction. This is systematically and consistently formulated in an effective field theory called  chiral perturbation theory (ChPT)~\cite{Weinberg:1978kz,Gasser:1983yg,Gasser:1984gg,Gasser:1987rb,Bernard:1995dp,Pich:1995bw,Bernard:2007zu,Scherer:2009bt}. At present only a few LQCD simulations have been preformed directly at the physical light quark masses~\cite{Durr:2008zz,Aoki:2009ix}, while most  calculations still require some kind of chiral extrapolation, often introducing sizable uncertainties to the final results.

LQCD simulations, by definition, are performed in a hypercube with its volume as a simulation parameter. The volume has to be large enough such that 
physics in a finite hypercube is  approximately the same as that in infinite space-time. To have effects of this origin under control, a rule of thumb\footnote{We will see that for the
ground-state octet baryon masses studied in this work, this requirement depends on the value of $m_\pi$: at fixed $m_\pi L$ the larger the $m_\pi$, the larger the finite-volume corrections, as previously noted in Ref.~\cite{Beane:2011pc}} is that $m_\pi L$ should be larger than $\sim4$. In the $p$-regime (where $m_\pi L\gg1$), it was first suggested by Gasser and Leutwyler that one could use ChPT to evaluate finite-volume corrections~\cite{Gasser:1986vb,Gasser:1987zq}. An alternative approach is the L\"uscher formula~\cite{Luscher:1985dn} and its resummed version (for a recent reference in the context of the nucleon mass, see Ref.~\cite{Colangelo:2010ba}). 

In recent years, it is found that three-flavor ($u$, $d$, and $s$) ChPT at next-to-leading order (NLO) has difficulties to accommodate recent LQCD results, particularly in the one-baryon sector.~\footnote{For an update on the present situation in the mesonic sector, see Ref.~\cite{Bernard:2010ex}.}  In the case of light hadron masses, it was shown that  NLO heavy baryon (HB) ChPT cannot describe the latest LHP~\cite{WalkerLoud:2008bp} and PACS-CS~\cite{Ishikawa:2009vc}  lattice data. On the other hand, covariant baryon ChPT supplemented with the extended-on-mass-shell (EOMS) scheme~\cite{MartinCamalich:2010fp} and ChPT regularized by a cutoff (long range regularization)~\cite{Young:2009zb} are shown to be able to describe much better the same lattice data.

In terms of chiral extrapolation, it seems that the advantage of  covariant ChPT over HBChPT in the one-baryon sector has been established, particularly in the three-flavor case. However, a detailed study of  finite-volume effects using three-flavor baryon ChPT is still missing. To perform such a study, it is advantageous to have LQCD simulations performed with the same setup except for the lattice size (volume). Such LQCD results have recently been provided by the NPLQCD collaboration~\cite{Beane:2011pc}, where simulations are performed with $n_f=2+1$ clover fermions in four lattice volumes, with spatial extent $L\sim2.0$, $2.5$, $3.0$, and $3.9$ fm, using an anisotropic lattice spacing of $b_s\sim0.123$ fm in the spatial direction and $b_t=b_s/3.5$ in the time direction, and at a pion mass of $m_\pi\approx390$ MeV. With these results, we can now perform a detailed study of finite-volume effects on the masses of ground-state octet baryons using both the covariant formulation of ChPT and its non-relativistic counterpart (HBChPT). 

In Ref.~\cite{Ali Khan:2003cu}, a covariant formulation of ChPT using the infrared scheme~\cite{Becher:1999he} up to next-to-next-to-leading order (NNLO) was used to study the dependence of nucleon masses on the lattice size $L$ in the $n_f=2$ case. The authors concluded that NNLO relativistic ChPT can describe well the finite-volume effects of nucleon masses. In this work, however, contributions of the intermediate $\Delta(1232)$ were not considered. In Ref.~\cite{Procura:2006bj}, the effects of the intermediate $\Delta(1232)$ were studied and the authors pointed out that its effects are important but can be encoded into the relevant low-energy constants and therefore explicit inclusion of the $\Delta(1232)$ into NNLO ChPT in the infrared scheme is not necessary.\footnote{The additional low-energy constant $e^{(3)}_1$, introduced to compensate the logarithmic dependence on the renormalization scale, makes a direct comparison of their NLO results with those presented in this work difficult.}   

In Ref.~\cite{Beane:2011pc}, on the other hand, it was shown that three-flavor HBChPT at NLO can describe the observed volume dependences reasonably well and NLO HBChPT with the decuplet states integrated out does not provide a reliable description of the finite-volume effects at a pion mass of 390 MeV. We will show that indeed at NLO virtual decuplet baryons play a more important role in HBChPT than in covariant ChPT. In fact, in HBChPT much of the observed finite-volume effects can only be explained when contributions of virtual decuplet baryons are taken into account. This is because the virtual octet baryon contributions to  finite-volume corrections  are larger in covariant ChPT than   in HBChPT, while the contributions of intermediate decuplet baryons are smaller, because of  relativistic effects.

This paper is organized as follows. In Section 2, we calculate the finite-volume effects on the masses of the ground-state octet baryons in the covariant formulation of baryon ChPT.  In Section 3, we compare finite-volume effects predicted by covariant baryon ChPT and those by HBChPT and study the latest NPLQCD results.  We conclude in Section 4.

\section{Finite-volume corrections to ground-state octet baryon masses}
\begin{figure}[t]
\centerline{\includegraphics[scale=0.4]{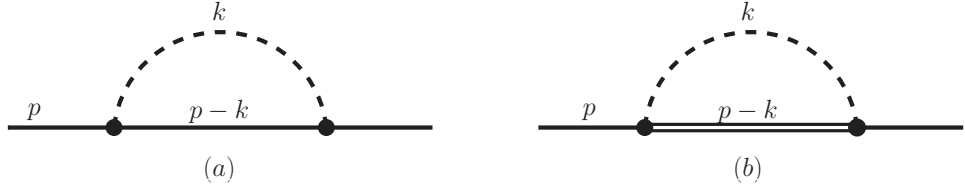}}
\caption{Feynman diagrams contributing to finite-volume effects on the masses of the ground-state octet baryons up to next-to-leading-order.
Solid lines denote octet baryons, solid double lines decuplet baryons, and dashed lines represent pseudoscalar mesons. \label{fig:diagrams}}
\end{figure}

In Ref.~\cite{MartinCamalich:2010fp}, the lowest-lying octet and decuplet baryon masses are calculated in a covariant formulation of
ChPT supplemented with the 
Extended-on-Mass-Shell (EOMS) renormalization scheme~\cite{Fuchs:2003qc,Geng:2009hh} to conserve proper chiral power counting.\footnote{In this work both EOMS ChPT and covariant ChPT will be used to refer to ChPT supplemented with the EOMS prescription,  unless otherwise explicitly specified.} It is shown that at NLO
 covariant ChPT can better describe the LQCD simulations compared to HBChPT.\footnote{In the study of light quark mass dependence of the $D$ and $D_s$ decay constants, it is also observed that covariant ChPT does a better
job compared to the heavy-meson ChPT~\cite{Geng:2010df}.} We will not repeat the same calculation here and refer interested readers to Ref.~\cite{MartinCamalich:2010fp} for the details. In this work we concentrate on finite-volume corrections and spell out details of the calculation which have been skipped in Ref.~\cite{MartinCamalich:2010fp}. 

Physically, finite-volume corrections can be easily understood: Because of the existence of space-time boundaries, the allowed momenta of virtual particles become discretized, i.e, one has to replace a momentum integral of the form $\int\limits_{-\infty}^{\infty} d k$ by an infinite sum of discretized momenta$\sum\limits_{n=-\infty}^\infty \left(\frac{2\pi}{L}\right) n$ (assuming periodical
boundary conditions). In LQCD simulations of zero-temperature physics the temporal extent is generally larger than the spacial extent such that the integral in the temporal dimension can be treated as if it extends from $-\infty$ and $\infty$. As a result, only the integral in the spacial dimensions  should be replaced by an infinite sum.
Obviously only loop diagrams are affected by the existence of space-time boundaries.  In studying the
octet baryon masses up to NLO,  only two such loop diagrams contribute and are shown in Fig.~\ref{fig:diagrams}.

To calculate the finite-volume corrections originating from these loop diagrams, one simply follows the conventional way of calculating Feynman diagrams and needs only to pay attention to the fact that now the temporal and the spacial dimensions must be treated differently. Diagrams (a) and (b) of Fig.~\ref{fig:diagrams} yield, generically,
\begin{equation}\label{eq:GN}
G_N=i\int \frac{d^4 k}{(2\pi)^4}\frac{\slashed{k}(\slashed{k}-\slashed{p}+M_0)\slashed{k}}{(k^2-m_M^2+i\epsilon)((p-k)^2-M_0^2+i\epsilon)},
\end{equation}
\begin{equation}\label{eq:GD}
G_D=i\int \frac{d^4 k}{(2\pi)^4}\frac{\gamma^{m a e}(p-k)^m k^e S^{ab}(p-k)\gamma^{n b f}(p-k)^n k^f}{(k^2-m_M^2+i\epsilon)((p-k)^2-M_D^2+i\epsilon)}
\end{equation}
with $\gamma^{mae}=\frac{1}{2}(\gamma^m\gamma^a\gamma^e-\gamma^e\gamma^a\gamma^m)$,
$\gamma^{nbf}=\frac{1}{2}(\gamma^n\gamma^b\gamma^f-\gamma^f\gamma^b\gamma^n)$, and
$S^{ab}(p)=-(\slashed{p}+M_D)(\eta^{ab}-\frac{1}{D-1}\gamma^a\gamma^b-\frac{1}{D-1}\frac{\gamma^a p^b-\gamma^b p^a}{M_D}-\frac{D-2}{D-1}\frac{p^a p^b}{M_D^2})$~\cite{Geng:2008bm}. In Eqs.~(\ref{eq:GN},\ref{eq:GD}), $M_0$ and $M_D$ are the
octet and decupet baryon masses at the chiral limit, and $m_M$ is the mass of a Nambu-Goldstone boson. As in Ref.~\cite{MartinCamalich:2010fp}, we have adopted the ``consistent'' coupling scheme advocated by Pascalutsa et al.~\cite{Pascalutsa:2000kd,Pascalutsa:2005nd} to describe the interactions between octet and decuplet baryons. In infinite space-time, the above integrals
have been calculated in Ref.~\cite{MartinCamalich:2010fp} and the results can be found there. As explained in
Ref.~\cite{MartinCamalich:2010fp}, the above loop functions contain power-counting-breaking (PCB) terms and therefore additional steps need to be taken to conserve a proper chiral power-counting scheme. Among the different approaches, the EOMS scheme has been shown to be superior to the heavy-baryon or infrared approaches (see Refs.~\cite{Geng:2008mf,Geng:2009hh,Geng:2010yc,Pascalutsa:2011fp} for an in-depth discussion on this topic). 

In the following, $G_{N/D}^\mathrm{(EOMS)}$ in the covariant framework represent the loop functions in which PCB terms have been removed using the EOMS prescription. Instead of calculating the integrals, Eqs.~(\ref{eq:GN},\ref{eq:GD}), in a finite hypercube, we calculate the following differences:
\begin{equation}
\delta G_N= G_N(L)-G_N(\infty),
\end{equation}
\begin{equation}
\delta G_D=G_D(L)-G_D(\infty),
\end{equation}
where $G_{N/D}(L)$ and $G_{N/D}(\infty)$ denote the integrals calculated in a finite hypercube and in infinite space-time. These quantities have several features that make calculations more feasible than a direct computation of $G_{N/D}(L)$.  First, because $G_{N/D}(L)$ and $G_{N/D}(\infty)$ have the same ultraviolet behavior, $\delta G_{N/D}$ are finite and can therefore be calculated in four dimensions. Second, the unwelcome PCB terms appearing in a covariant baryon ChPT calculation are absent because they emerge from short-distance physics while such short-distance properties are the same in $G_{N/D}(L)$ and $G_{N/D}(\infty)$. As a result,  PCB terms vanish  in the differences $\delta G_{N/D}$ and no power-counting-restoration schemes, such as EOMS or IR, are needed to calculate $\delta G_{N/D}$. 

For
$G_N$, one has
\begin{equation}\label{eq:Gn1}
G_N=i\int\limits^1_0 dx\int \frac{d^4 k}{(2\pi)^4}\frac{\slashed{k}(\slashed{k}-\slashed{p}+M_0)\slashed{k}}{((k-px)^2-\mathcal{M}_N^2)^2}
=i\int\limits^1_0 dx\int \frac{d^4k}{(2\pi)^4} \frac{ \slashed{k}(k^2-2k\cdot p)+2 k^2 M_0}{((k-px)^2-\mathcal{M}_N^2)^2},
\end{equation}
where  $\mathcal{M}^2_N=x^2 M_0^2+(1-x)m_M^2-i\epsilon$.
Calculating the integral [Eq.~(\ref{eq:Gn1})] in a finite hypercube requires treating the temporal and spacial dimensions differently.
We choose to work in the baryon rest frame, i.e., $p^\mu=(M_0,\vec{0})$. In this frame,\begin{equation}
G_N=i\int\limits^1_0 dx\int \frac{d k_0}{2\pi} \int \frac{d\vec{k}}{(2\pi)^3} \frac{ (\gamma^0 k_0-\vec{\gamma}\cdot\vec{k})(k_0^2-\vec{k}^2-2k_0 M_0)+2 (k_0^2-\vec{k}^2) M_0}{((k_0-x M_0)^2-\vec{k}^2-\mathcal{M}_N^2)^2}.
\end{equation}
This can be easily calculated by performing a shift in $k_0$ ($k_0\rightarrow k_0'+x M_0$), Wick rotating $k'_0$ ($k'_0\rightarrow ik'_0$), and then performing the integration over $k'_0$. The result is:
\begin{equation}
G_N=\int\limits^1_0 dx\int\frac{d\vec{k}}{(2\pi)^3} \left[\frac{1}{2}M_0(2x+1)\left(\frac{1}{\vec{k}^2+\mathcal{M}_N^2}\right)^{1/2}-\frac{1}{4}M_0(M_0^2 x^3+\mathcal{M}_N^2(x+2))\left(\frac{1}{\vec{k}^2+\mathcal{M}_N^2}\right)^{3/2}\right].
\end{equation}
From this and utilizing the master formula as provided, e.g., in Ref.~\cite{Beane:2004tw}, one can easily obtain
\begin{equation}\label{eq:DeltaGN}
\delta G_N=\int\limits^1_0 dx \left[\frac{1}{2}M_0(2x+1)\delta_{1/2}(\mathcal{M}^2_N)-\frac{1}{4}M_0(M_0^2 x^3+\mathcal{M}^2_N(x+2))\delta_{3/2}(\mathcal{M}^2_N)\right],
\end{equation}
where 
\begin{equation}
\delta_r (\mathcal{M}^2)=
\frac{2^{-1/2-r}(\sqrt{\mathcal{M}^2})^{3-2r}}{\pi^{3/2}\Gamma(r)}\sum_{\vec{n}\ne0}(L\sqrt{\mathcal{M}^2}|\vec{n}|)^{-3/2+r}K_{3/2-r}(L\sqrt{\mathcal{M}^2}|\vec{n}|),
\end{equation}
where $K_n(z)$ is the modified Bessel function of the second kind, and $\sum\limits_{\vec{n}\ne0}\equiv\sum\limits^{\infty}_{n_x=-\infty}\sum\limits^{\infty}_{n_y=-\infty}\sum\limits^{\infty}_{n_z=-\infty}(1-\delta(|\vec{n}|,0))$ with $\vec{n}=(n_x,n_y,n_z)$.

In a similar way, one can calculate $\delta G_D$:
\begin{equation}\label{eq:DeltaGD}
\delta G_D=\int\limits^1_0 dx\left[\frac{M_0^2 (M_0(1-x)+M_D)}{6 M_D^2}\delta_{1/2}(\mathcal{M}^2_D)
-\frac{M_0^2(M_0(1-x)+M_D)\mathcal{M}_D^2}{6 M_D^2}\delta_{3/2}(\mathcal{M}^2_D)\right],
\end{equation}
where $\mathcal{M}^2_D= x^2 M_0^2 - x(M_0^2-M_D^2)+(1-x) m_M^2 - i\epsilon$.
 
The corresponding results in HBChPT can be found in Ref.~\cite{Beane:2004tw} and are given below
\begin{equation}
\delta G_{N/D}^\mathrm{(HB)}=C_{N/D}\int\limits^\infty_0 d\lambda \beta_\Delta \sum\limits_{\vec{n}\ne0}\left[(L|\vec{n}|)^{-1} K_1(L\beta_\Delta|\vec{n}|)-\beta_\Delta K_0(L\beta_\Delta |\vec{n}|)\right],
\end{equation}
where $\beta_\Delta^2\equiv\lambda^2+2\lambda\Delta+m_M^2$ with $\Delta=M_D-M_0$ for $\delta G_D^\mathrm{(HB)}$ and $\Delta=0$ for $\delta G_N^\mathrm{(HB)}$, $C_N=-4$, and $C_D=-16/6$. 

Numerically one can easily check that $\delta G_{N/D}$ [Eqs.~(\ref{eq:DeltaGN},\ref{eq:DeltaGD})] turn out to be identical to $\delta G_{N/D}^{(\mathrm{HB})}$ in the limit of
$M_0\rightarrow\infty$, which confirms our statement that there are no PCB terms in $\delta G_{N/D}$.  Furthermore, we have noted that changing the integration region in Eqs.~(\ref{eq:DeltaGN},\ref{eq:DeltaGD}) from $\int\limits^1_0$ to $\int\limits^\infty_0$, which corresponds to the infrared prescription~\cite{Ali Khan:2003cu}, has negligible effects on the numerical
results of those integrals.

\begin{figure}[t]
\centering
\subfigure{\includegraphics[angle=270,scale=0.32]{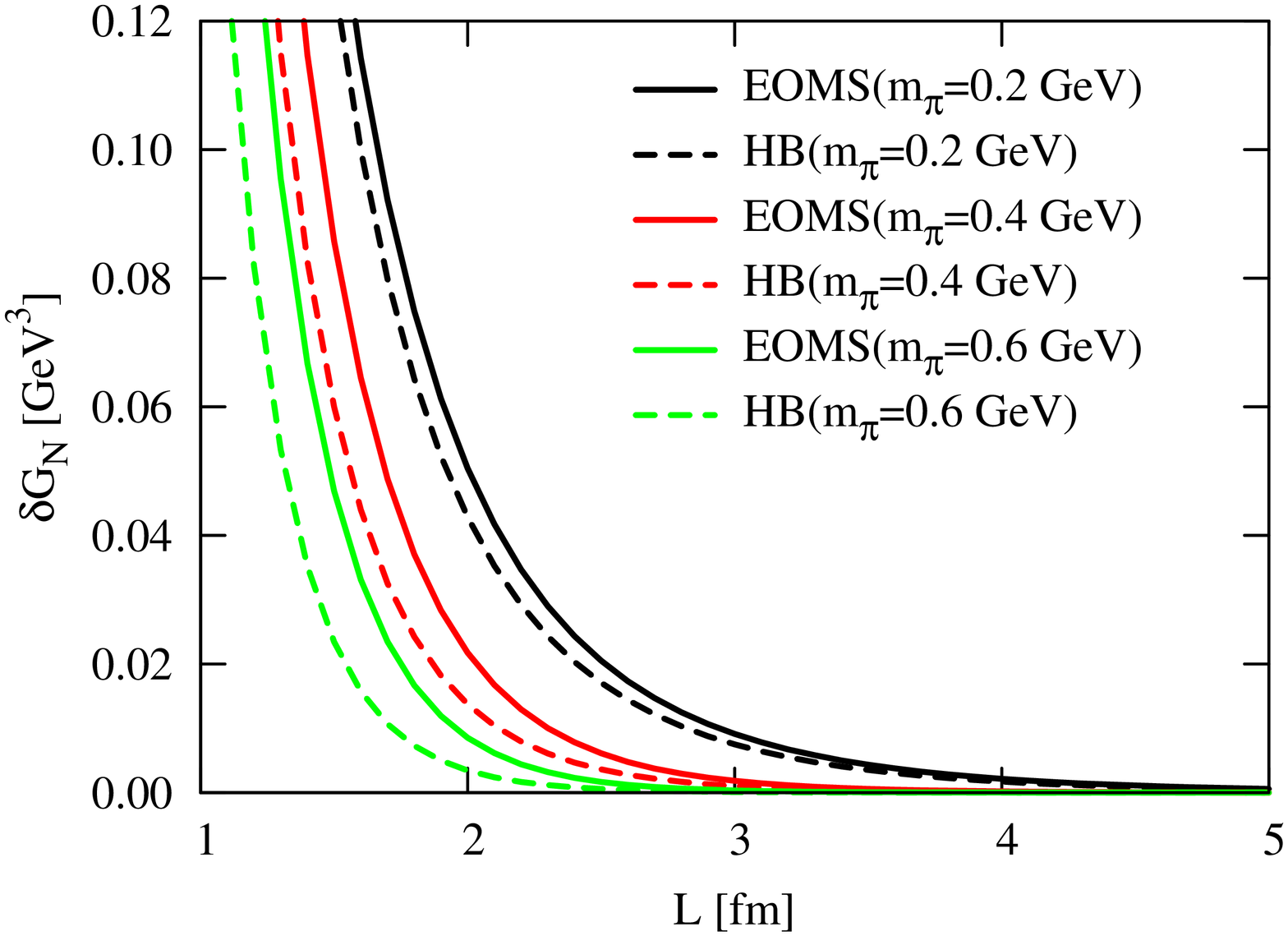}}
\subfigure{\includegraphics[angle=270,scale=0.32]{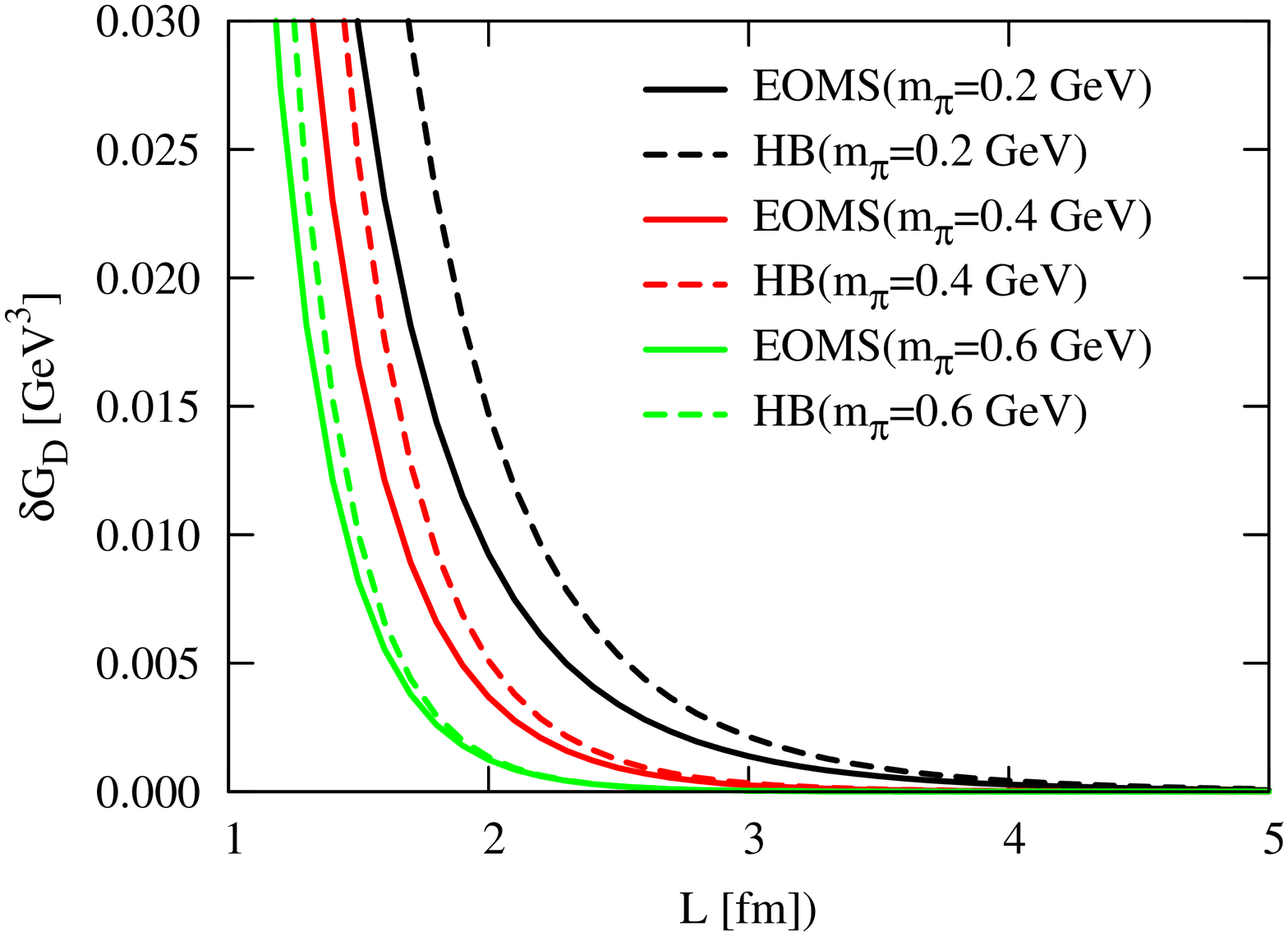}}
\caption{Finite-volume corrections, $\delta G_N$ (left) and $\delta G_D$ (right), as functions of lattice size $L$ for $m_\pi=0.2$, $0.4$ and $0.6$ GeV.  In the evaluation of $\delta G_{D/N}$  in covariant ChPT $M_0=0.8$ GeV. The decuplet-octet mass gap is set to be $\Delta=0.231$ GeV.\label{fig:deltaND}}
\end{figure}
\section{Results and discussion}
\subsection{Finite-volume corrections: HB vs. EOMS ChPT}
Before looking at the NPLQCD results, it is instructive to compare the finite-volume corrections computed in 
covariant ChPT and  HBChPT.   In Fig.~\ref{fig:deltaND}, finite-volume corrections $\delta G_{N/D}$ for different $m_\pi$ are plotted as functions of the lattice size $L$. The well-known rapid decrease of finite-volume corrections with increasing $L$ is clearly seen. The finite-volume corrections can be parameterized as a linear function of $e^{-m_\pi L}/(m_\pi L)$, as indicated by explicit calculations for large $L$ with both ChPT and the L\"uscher method.\footnote{To factor out such an exponential dependence, in the rest of this work, we will plot relevant quantities as functions of $e^{-m_\pi L}/(m_\pi L)$.} It is clear that for $\delta G_N$  at fixed $m_\pi$ and $L$ the covariant results are larger than the HB results. The differences become smaller as $m_\pi$ decreases. On the other hand, for $\delta G_D$ the covariant results are smaller than the HB results and the differences become larger as $m_\pi$ decreases.  It should be stressed that the enhancement of virtual octet contributions and reduction of virtual decuplet contributions in EOMS compared to HB ChPT have important consequences for extraction of the decuplet-octet axial couplings from finite-volume dependence of octet baryon masses, as we will show in the following subsection.

\begin{figure}[t]
\centering
\subfigure{\includegraphics[angle=270,scale=0.32]{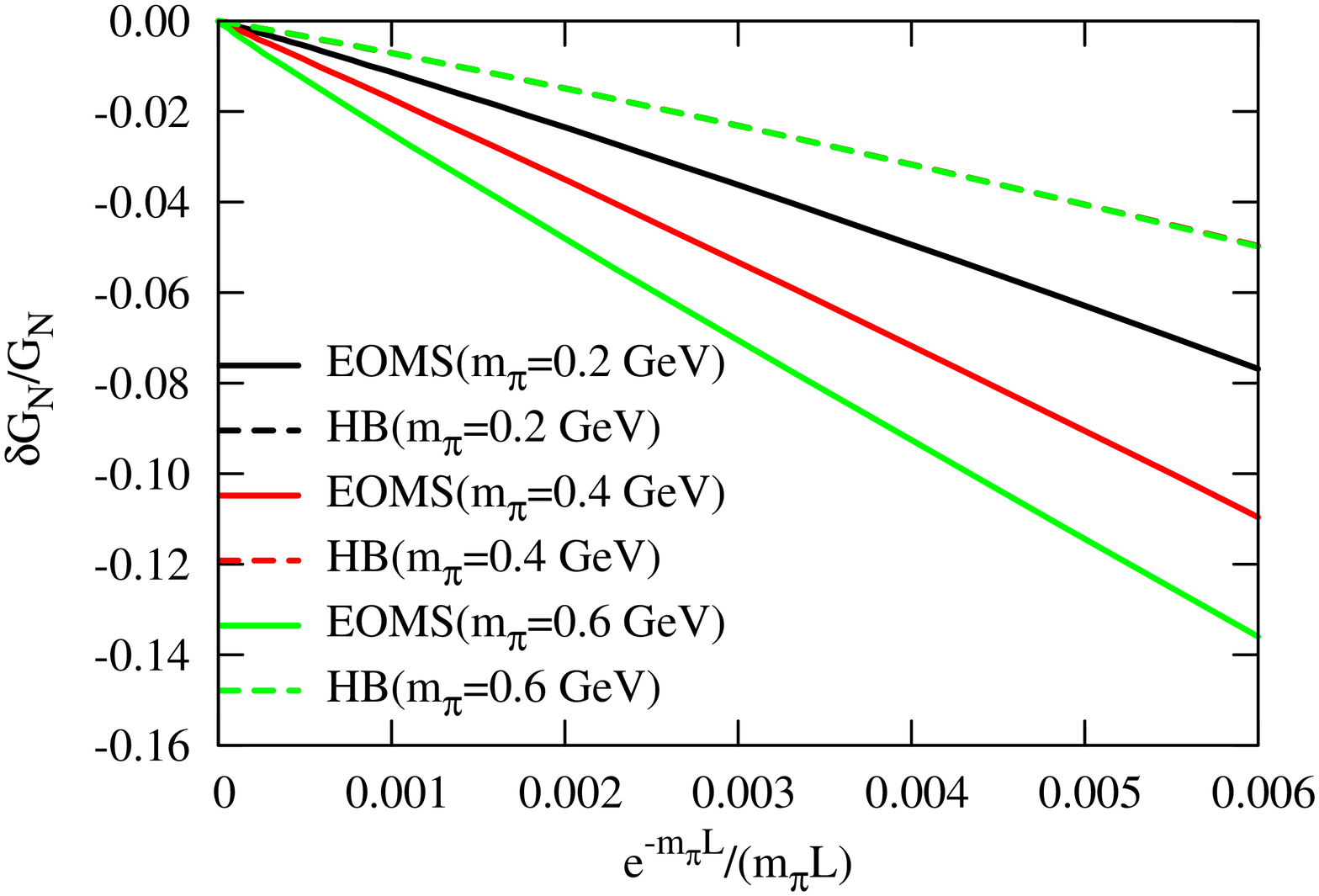}}
\subfigure{\includegraphics[angle=270,scale=0.32]{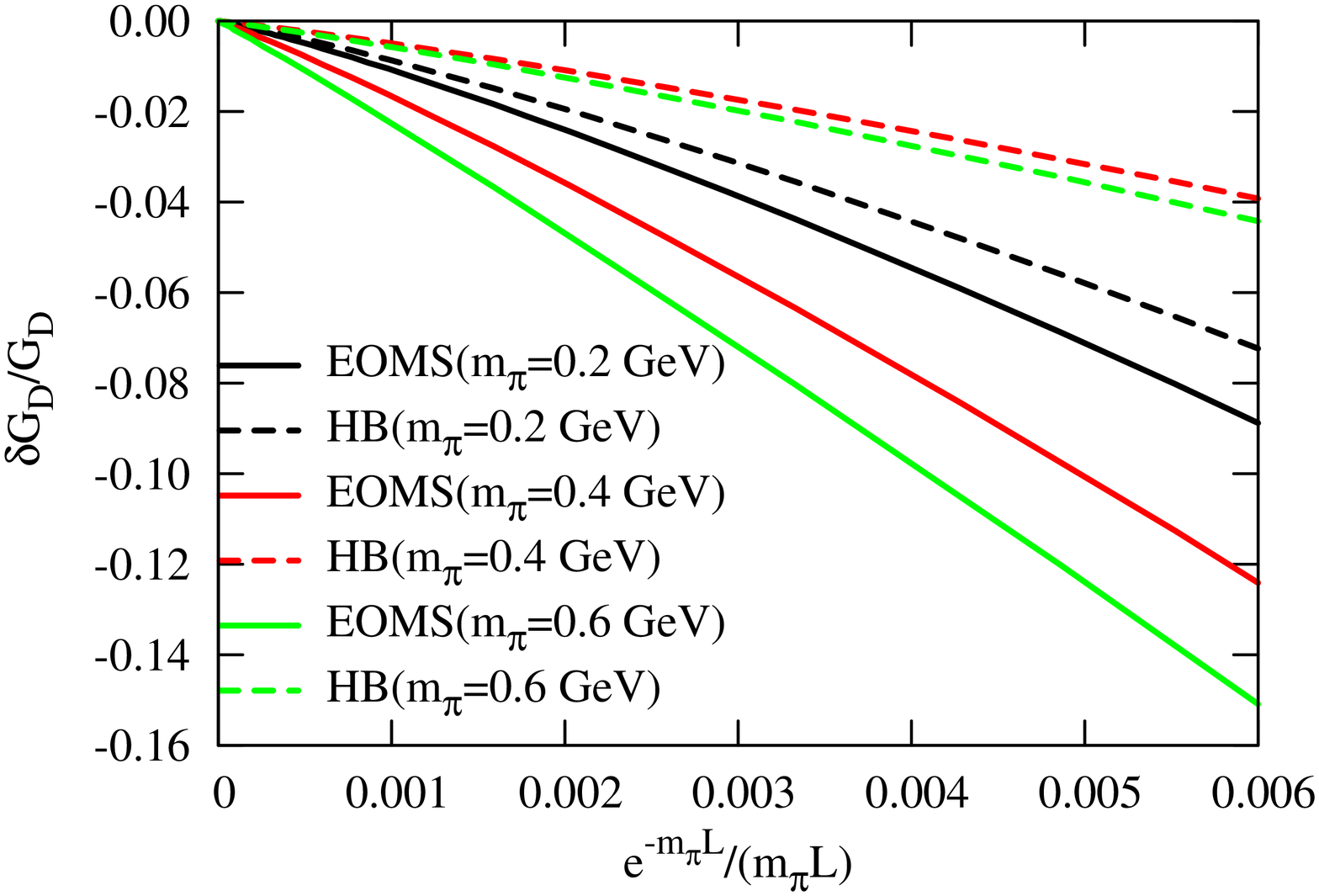}}
\caption{Finite-volume effects, $\delta G_N/G_N$ (left) and $\delta G_D/G_D$ (right), as functions of $e^{-m_\pi L}/(m_\pi L)$ for $m_\pi=0.2$, $0.4$ and $0.6$ GeV.  In the evaluation of $\delta G_{D/N}$ ($G_{D/N}$) in EOMS ChPT $M_0=0.8$ GeV. The decuplet-octet mass gap is set to be $\Delta=0.231$ GeV.\label{fig:deltaNDratio}}
\end{figure}

In ChPT, where finite-volume corrections are calculated order by order from loop diagrams, the ratios of $\delta G_{N/D}/G_{N/D}$ demonstrate better the differences between various ways of calculating the loop diagrams, e.g., HB vs. EOMS. In Fig.~\ref{fig:deltaNDratio}, we plot $\delta G_{N/D}/G_{N/D}$  as  functions of
$e^{-m_\pi L}/(m_\pi L)$ for $m_\pi=0.2$, $0.4$, and $0.6$ GeV, respectively. The finite-volume effects  decrease almost linearly in both cases, reflecting the fact that to a large extent they can be parameterized as a linear function in terms of $e^{-m_\pi L}/(m_\pi L)$. Such a feature has been employed in Ref.~\cite{Beane:2011pc} to perform a phenomenological fit to their data. There are, however, some subtle differences between the EOMS and HB results,  which one cannot see easily from Fig.~\ref{fig:deltaND}. At fixed $m_\pi L$, in the covariant case the larger the $m_\pi$, the larger the finite-volume effects.  On the other hand, in HBChPT $\delta G_N/G_N$ depends only on $m_\pi L$,\footnote{This can be immediately understood by looking at the corresponding analytical results:
\begin{eqnarray}
G_N^{\mathrm{(HB)}}&=&-2\pi m_\pi^3,\\
\delta G_N^{\mathrm{(HB)}}&=& 2\pi m_\pi^3\sum\limits_{\vec{n}\ne0}(Lm_\pi |\vec{n}|)^{-1}\exp(-m_\pi L|\vec{n}|).
\end{eqnarray}}
while the dependences of $\delta G_D/G_D$  on $m_\pi$  are clearly different from its covariant counterpart.

\subsection{Study of  the NPLQCD results}
\begin{figure}[t]
\centerline{\includegraphics[scale=0.4]{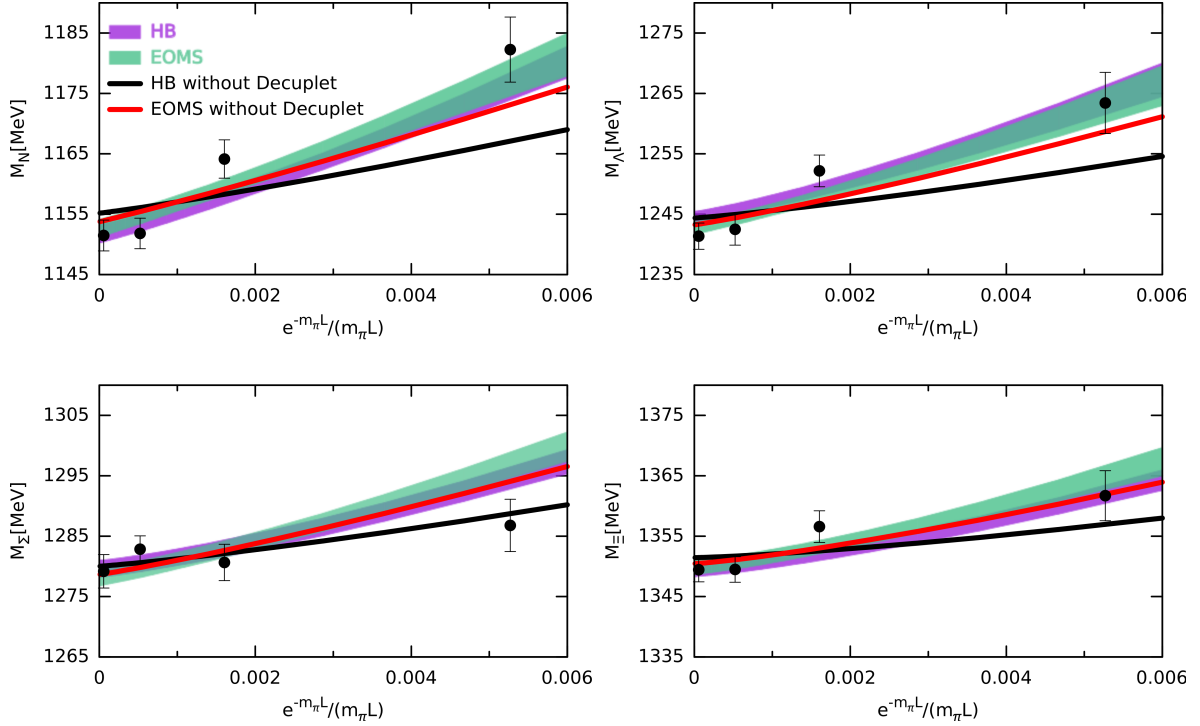}}
\caption{The NPLQCD octet mass data~\cite{Beane:2011pc} fitted with NLO covariant ChPT and HBChPT. The bands are the full results at the 68\% confidence level and the solid (dashed) lines
are the best fits with $C=0$.\label{fig:res}}
\end{figure}

Now let us turn to the NPLQCD data. The lattice results in units of temporary lattice spacing are given in Table I of Ref.~\cite{Beane:2011pc}, which can be translated into physical units using  $b_t=b_s/3.5$ with $b_s=0.1227\pm0.0008$ fm. We have chosen to fit the masses with physical units, which is equivalent to fitting the masses with lattice units. Of course, to get the final physical results one must take into account also the uncertainties due to the determination of the lattice spacing. This, however, does not affect the quality of our fits.

The octet baryon masses at NLO covariant baryon ChPT in infinite space-time have the following form:
\begin{equation}
M^{(3)}_{\mathcal{B}}=m_0-\sum_{\phi=\pi,K}\xi^{(a)}_{\mathcal{B},\phi}m_\phi^2+\frac{1}{(4\pi F_\phi)^2}\sum_{\substack{\phi=\pi,K,\eta\\ \alpha=b,c}}\xi^{(\alpha)}_{\mathcal{B},\phi} H_X^{(\alpha)}(m_\phi)\label{Eq:ResGen},
\end{equation}
where $\xi^{(a)}_{\mathcal{B},\phi}$ and $\xi^{(\alpha)}_{\mathcal{B},\phi}$ are tabulated in Table V of Ref.~\cite{MartinCamalich:2010fp},
the loop functions $H_X^{(\alpha)}$ are given in Eqs.~(A3, A4) in the same reference, which contain PCB terms that have to be removed by the EOMS prescription as explained in detail in Ref.~\cite{MartinCamalich:2010fp}. To calculate the baryon masses in
a finite hypercube, one simply replaces the loop functions $H_X^{(\alpha)}$ with their counterparts calculated in a finite hypercube as provided in the previous section, i.e.,
\begin{eqnarray}
H_X^{(b)}&\rightarrow& H_X^{(b)}+\frac{1}{2}\delta G_N(L),\\
H_X^{(c)}&\rightarrow& H_X^{(c)}+\frac{3}{4}\delta G_D(L).
\end{eqnarray}

To fit the octet baryon masses in the covariant ChPT one has four low-energy constants (LECs) to determine: $b_0$, $b_D$, $b_F$, $M_0$. The other parameters are given the same values as those used in Ref.~\cite{MartinCamalich:2010fp} , i.e., $D=0.8$, $F=0.46$, $\mu=1$ GeV, 
$F_\pi=F_K=F_\eta=1.17 f_\pi$ with $f_\pi=92.4$ MeV.\footnote{It should be noted that in the present study, SU(3) breaking effects are entirely induced by the masses of the pseudoscalar mesons, while for the couplings we have not introduced any explicit SU(3) symmetry breaking, because they can not be determined by the NPLQCD data.}  Since in the present case all the lattice data are obtained at the same pion mass, one could not distinguish $b_0$ and $M_0$ and therefore we have taken $M_0=0.8$ GeV, as suggested by the covariant ChPT study of  the PACS and LHP  data~\cite{MartinCamalich:2010fp,MartinCamalich:2010zz}. A moderate variation of $M_0$ from its central value by, e.g., 0.2 GeV, does not change the results in any appreciable way. It should be noted that $M_0$ does not enter  loop calculations in HBChPT. For the decuplet-octet mass gap, we take the average value $\Delta=0.231$ GeV. Using the mass gap between the nucleon and the $\Delta(1232)$, 0.291GeV, for $\Delta$ has minor effects on our fits. Furthermore, to 
study the effects of the intermediate decuplet baryons we allow $C$, the decuplet-octet axial coupling, to vary. At the end, we have four parameters to fit 16 lattice data.

The fitted results using both covariant ChPT and HBChPT are shown in Fig.~\ref{fig:res}. Both methods provide a reasonable fit to the lattice data with similar quality, compatible with the phenomenological fit performed in Ref.~\cite{Beane:2011pc}, though  with less parameters (four vs. eight). On the other hand, for the $\Sigma$, ChPT seems to predict a larger finite-volume dependence than suggested by the lattice data. The values of the LECs corresponding to our best fits are given in Table I.  Two things are noteworthy. First,
the values of $b_0$, $b_D$, and $b_F$ are close to those obtained in Ref.~\cite{MartinCamalich:2010fp}.  Second, the central value of $C$ from the best fit in EOMS ChPT is only about $3/4$ of the value we used in Ref.~\cite{MartinCamalich:2010fp}, which is fixed from the $\Delta$ decay width. Because $C$ should be understood as an average  of all the
decuplet-octet axial couplings, such a value, taking into account its uncertainty, is not out of the range of our expectations. On the other hand, the value of $C$ fixed from the HPChPT fit is slightly larger, $0.94\pm0.09$. This is larger than the value used in most HB calculations, e.g., $C=g_{\Delta N}/2=0.7$~\cite{Beane:2004tw} and $C=0.76$~\cite{Young:2009zb}. Both results are, however, roughly consistent with each other, keeping in mind that NNLO ChPT contributions could be as large as 30\%  of those of the NLO.

\begin{table*}[t]
      \renewcommand{\arraystretch}{1.6}
     \setlength{\tabcolsep}{0.2cm}
     \centering
     \caption{\label{table:par}Values of the low-energy constants from the best fit to the NPLQD data with $\chi^2/\mathrm{d.o.f.}\approx1.6$. }
     \begin{tabular}{c|cccc}
     \hline\hline
            & $b_0$ & $b_D$ & $b_F$ &  $C$   \\ \hline
  EOMS& $-0.81\pm0.08$ & $0.12\pm0.04$ & $-0.46\pm0.04$ & $0.73\pm0.23$ \\
  HB    &  $ -1.85\pm0.20$ & $0.61\pm0.11$ & $-1.01\pm0.09$ & $0.94\pm0.09$\\
 \hline\hline
    \end{tabular} 
\end{table*}
\begin{table*}[b]
      \renewcommand{\arraystretch}{1.6}
     \setlength{\tabcolsep}{0.2cm}
     \centering
     \caption{\label{table:par}Extrapolated octet baryon masses in the infinite lattice size limit. }
     \begin{tabular}{c|cccc}
     \hline\hline
            &  EOMS ChPT & HBChPT & SU2 HBChPT~\cite{Beane:2011pc}   \\ \hline
 $M_N$& $1152.7\pm1.4$ & $1151.7\pm1.5$ & $1151.3\pm1.1$\\
  $M_\Lambda$& $1242.7\pm1.0$ & $1244.5\pm1.0$ & $1241.9\pm0.8$ \\
  $M_\Sigma$ & $1278.2\pm1.4$ & $1279.6\pm1.4$ & $1280.3\pm1.0$\\
   $M_\Xi$ & $1349.8\pm1.2$ & $1349.5\pm1.2$ & $1349.6\pm0.7$\\  
   \hline\hline
    \end{tabular} 
\end{table*}

The effects of the virtual decuplet baryons can be best seen by fitting the NPLQCD data with $C=0$. The corresponding results are shown  by the solid (dashed) lines in Fig.~\ref{fig:res}. It is clear that in the fit the octet-decuplet transition plays a larger role in HB than in covariant ChPT. In fact, in HBChPT virtual decuplet baryons play an even larger role than
those of virtual octet baryons, which seems to be a bit unnatural (for a relevant discussion, see, e.g., Ref.~\cite{Geng:2009hh}). In Ref.~\cite{Beane:2011pc}, it was concluded that the decuplet contributions must be taken into account. Our studies show that this is indeed the case, but more so in the HBChPT than in the covariant ChPT.  From the above discussion, we reach the same conclusion as Ref.~\cite{Beane:2011pc} that
extraction of the decuplet-octet axial couplings from the present lattice data by studying the volume dependence of
the octet baryons masses cannot be taken too seriously.

\begin{figure}[t]
\centerline{\includegraphics[scale=0.65,angle=270]{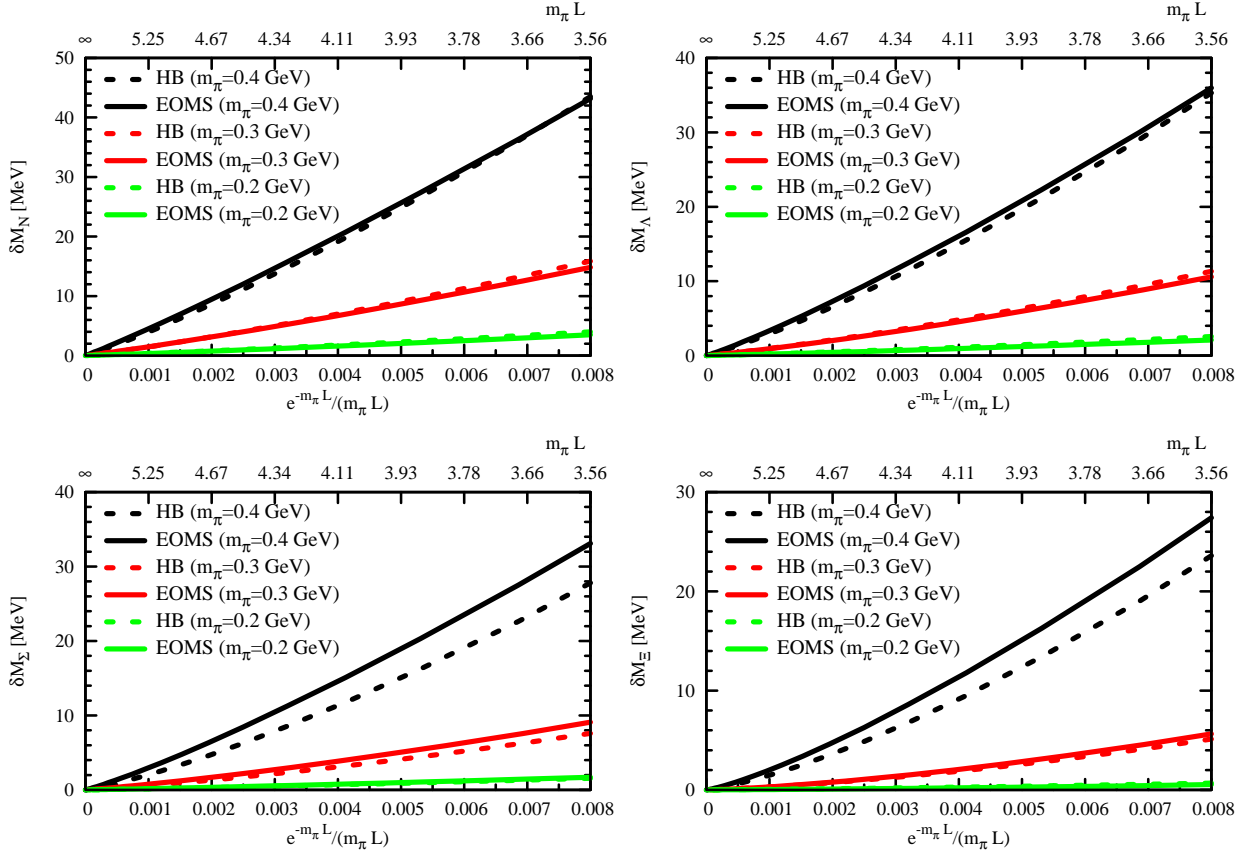}}
\caption{Predicted finite-volume corrections for the ground-state octet baryons using the best
fitted LECs determined from the NPLQCD data~\cite{Beane:2011pc}, where $m_s$ is assumed to have its physical value, and
$m_K$ ($m_\eta$) are related to $m_\pi$ through leading-order ChPT: $m_K^2=\frac{1}{2}(2B_0 m_s+ m_\pi^2)$, $m_\eta^2=\frac{1}{3}(4 m_K^2-m_\pi^2)$,
and $2 B_0 m_s=(2 m_K^2-m_\pi^2)|_\mathrm{phys}$.\label{fig:pre}}
\end{figure}

The extrapolated octet baryon masses in the limit of $L\rightarrow\infty$ are shown in Table II. For the sake of comparison, we
have tabulated the results of the NLO SU(2) HBChPT fit from Ref.~\cite{Beane:2011pc}. The uncertainties are
from the lattice data (systematic plus statistical) and we have ignored our systematic uncertainties and the uncertainties due to the lattice scale in converting the
lattice results from lattice units to physical units.\footnote{As can be seen  from Table III of Ref.~\cite{Beane:2011pc}, such uncertainties are far more important numerically. }  We note that the extrapolated masses using different
methods are roughly consistent with each other. This is mainly because the lattice simulations have been performed 
with a large volume, $L=3.9$ fm, where finite-volume corrections are almost zero and which strongly constrain the extrapolations.

In Fig.~\ref{fig:pre}, we show the predicted finite-volume corrections for different $m_\pi$ using the LECs given in Table I. The HB and covariant ChPT predictions are quite similar, except for the $\Sigma$ and $\Xi$, where  the difference could reach  $10-20$\% at a pion mass of 0.4 GeV and small $L$. Therefore, it can be concluded that at NLO one can use either HBChPT or covariant ChPT to describe finite-volume corrections to the octet baryon masses, keeping in mind that relativistic effects are large and one must be careful about interpretation of the extracted physical quantities.

\section{Summary and conclusions}
We have studied finite-volume effects on the octet baryon masses by analyzing the latest $n_f=2+1$ NPLQCD data
with a covariant formulation of baryon chiral perturbation theory and with heavy baryon chiral perturbation theory. It was shown that
although both approaches can describe the lattice data reasonably well, the underlying physics is different: Decuplet contributions
play a less important role in covariant ChPT than in HBChPT  at next-to-leading order because relativistic corrections enhance virtual octet contributions and reduce intermediate decuplet contributions. This makes it difficult to reliably extract the values of the 
decuplet-octet axial couplings from the volume dependences of the octet baryon masses. Simulations with a larger statistics, multiple volumes, and different pion masses will likely better serve such purposes.

We have shown that at fixed $m_\pi L$ ($\gg1$) finite-volume corrections become smaller for smaller $m_\pi$, as the NPLQCD collaboration
has pointed out. For the LQCD calculation of the nucleon mass,  $m_\pi L\ge4$ is needed to have an finite-volume correction at the order of 1\% at a pion mass of $\sim0.3$ GeV, while
at  a pion mass of $\sim0.2$ GeV, $m_\pi L\ge3$ is enough. For the other
octet baryons, the dependence of finite-volume corrections on $e^{-m_\pi L}/(m_\pi L)$ is weaker than that of the nucleon, provided that $m_s$ is close to its physical value.

It should be stressed that the fact that both HBChPT and covariant ChPT can describe finite-volume effects on the ground-state octet baryon masses does not mean that both approaches are capable of describing the pion-mass dependence of these quantities. In fact, it has been shown that at NLO covariant ChPT is more suitable for the purpose of chiral extrapolation.

At present, to go to higher chiral orders in the applications of three-flavor covariant baryon ChPT, one faces the problem of a large number of  poorly known low-energy constants, such as the decuplet-octet axial couplings. LQCD simulations with different quark masses and a sequence of different volumes help setting constraints on the values of these LECs as shown in the present work.

\section{Acknowledgements}
LSG acknowledges instructive discussions with Norbert Kaiser. This Work is supported in part by BMBF, the A.v. Humboldt foundation, the Fundamental Research Funds for the Central Universities, the National Natural
Science Foundation of China (Grant No. 11005007), and by the DFG Excellence Cluster ``Origin and Structure of the Universe."  MC acknowledges support from the MEC (contract FIS2006-03438), the EU Integrated Infrastructure Initiative Hadron Physics Project (contract RII3-CT-2004-506078) and the  Science and Technology Facilities Council (grant number ST/H004661/1).

\end{document}